\documentclass[11pt]{article}

\usepackage{SIGACTNews}
\usepackage{chicagor}

\usepackage{amsmath}
\usepackage{amsfonts}
\usepackage{amssymb}
\usepackage{wasysym}

\allowdisplaybreaks[2]

\newenvironment{prog}{\begin{array}[t]{@{}l@{}}}{\end{array}}

\newcommand{\thmcolon}{\hspace{-1ex}{\bf .}}

\newtheorem{THEOREM}{Theorem}%
\newtheorem{LEMMA}[THEOREM]{Lemma}
\newtheorem{COROLLARY}[THEOREM]{Corollary}
\newtheorem{PROPOSITION}[THEOREM]{Proposition}
\newtheorem{DEFINITION}[THEOREM]{Definition}
\newtheorem{CLAIM}[THEOREM]{Claim}
\newtheorem{EXAMPLE}[THEOREM]{Example}
\newtheorem{REMARK}[THEOREM]{Remark}

\newenvironment{theorem}{\begin{THEOREM} \thmcolon}%
                        {\end{THEOREM}}
\newenvironment{lemma}{\begin{LEMMA} \thmcolon  }%
                      {\end{LEMMA}}
                          {\end{COROLLARY}}
                            {\end{PROPOSITION}}
                            { \end{DEFINITION}}
                            {\end{CLAIM}}

                            { \wbox\end{EXAMPLE}}
\newenvironment{example*}{\begin{EXAMPLE} \thmcolon  \rm}%
                            {\end{EXAMPLE}}
                            { \wbox\end{REMARK}}
\newenvironment{remark*}{\begin{REMARK} \thmcolon  \rm}%
                            {\end{REMARK}}

\def\squareforqed{\hbox{\rlap{$\sqcap$}$\sqcup$}}
\def\wbox{\ifmmode\squareforqed\else{\unskip\nobreak\hfil
\penalty50\hskip1em\null\nobreak\hfil\squareforqed
\parfillskip=0pt\finalhyphendemerits=0\endgraf}\fi}
\newenvironment{proof}{\noindent{\it Proof.}}
                      {\wbox\vspace{0.1in}}

\newcommand{\nsat}{\not\models}

\newcommand{\Next}{\ocircle}
\newcommand{\GNext}{\ensuremath{\mathord{\mathbin{\settowidth{\dimen7}{\mbox{$\ocircle$}}%
              \makebox[0pt][l]{$\ocircle$}\makebox[\dimen7]{$\cdot$}}}}}
\newcommand{\Until}{\mathop{\mathcal{U}}{}}
\newcommand{\Sometimes}{{\lower 1pt\hbox{$\Diamond$}}}
\newcommand{\Always}{\Box}
\newcommand{\falsep}{\mathit{false}}
\newcommand{\truep}{\mathit{true}}

\newcommand{\cM}{\mathcal{M}}
\newcommand{\cMinf}{\mathcal{M}^{\mathit{inf}}}
\newcommand{\cMfin}{\mathcal{M}^{\mathit{fin}}}
\newcommand{\Ax}{\text{\textit{ax}}}
\newcommand{\AX}{\text{\textbf{AX}}}
\newcommand{\AXgen}{\AX^\mathit{gen}}
\newcommand{\AXinf}{\AX^\mathit{inf}}
\newcommand{\AXfin}{\AX^\mathit{fin}}
\newcommand{\AXcr}{\AX^\mathit{cr}}
\newcommand{\axiom}[1]{\mbox{#1}}
\newcommand{\FL}{\mathit{Cl}}
\newcommand{\At}{\mathit{At}}

\newenvironment{wideitemize}[1]
   {\begin{list}{$\bullet$}
                     {\setlength{\labelwidth}{#1}
                      \setlength{\leftmargin}{#1}}}
   {\end{list}}
\newenvironment{axiomlist}
   {\begin{wideitemize}{14ex}}
   {\end{wideitemize}}

\newcommand{\caret}{\textsc{CaRet}}
\newcommand{\call}{\mathit{call}}
\newcommand{\ret}{\mathit{ret}}
\newcommand{\inter}{\mathit{int}}

\usepackage{url}

\newcommand{\sat}{\models}
\newcommand{\rimp}{\Rightarrow}
\newcommand{\riff}{\Leftrightarrow}
\renewcommand{\phi}{\varphi}
\newcommand{\<}{\langle}
\renewcommand{\>}{\rangle}
\renewcommand{\emptyset}{\varnothing}

\newcommand{\COMMENTOUT}[1]{}

\title{SIGACT News Logic Column 11\\[2ex]
\textbf{The Finite and the Infinite in Temporal
Logic}\footnote{\copyright{} Riccardo Pucella, 2005. This version
differs slightly from that published in SIGACT News 36(1); it corrects
a number of typos in the semantics. Thanks to Claudia Zepeda for
pointing them out.}}
\author{Riccardo Pucella\\
Cornell University\\
Ithaca, NY 14853 USA\\
riccardo@cs.cornell.edu}
\date{}

\begin{document}

\SIGACTmaketitle

{\noindent\it\textbf{Note from the Editor:} I am always looking for
contributions. If you have any suggestion concerning the content of
the Logic Column, or even better, if you would like to contribute by
writing a survey or tutorial on your own work or topic related to your
area of interest, feel free to get in touch with me.\\[3ex]}

At the last TACAS in Barcelona, already almost a year ago, Alur,
Etessami, and Madhusudan \citeyear{r:alur04} introduced \caret, a
temporal logic framework for reasoning about programs with
nested procedure calls and returns. The details of the logic were
themselves interesting (I will return to them later), but a thought
struck me during the presentation, whether an axiomatization might
help understand the new temporal operators present in \caret{}. Thinking a
bit more about this question quickly led to further questions about
the notion of finiteness and infinity in temporal logic as it is used
in Computer Science. This examination of the properties of temporal
logic operators under finite and infinite interpretations is the topic
that I would like to discuss here. I will relate the discussion back
to \caret{} towards the end of the article, and derive a sound and
complete axiomatization for an important fragment of the logic.

Temporal logic is commonly used in Computer Science to reason about
temporal properties of state sequences
\cite{r:pnueli77,r:gabbay80}. Generally, these state sequences are the
states that arise during the execution of a program. Temporal logic
lets one write down properties such as ``an acquired lock is
eventually released'' or ``it is never the case that the value of such
variable is zero''. These kinds of properties become even more
important in concurrent programs, where properties such as ``every
process eventually executes its critical section'', or ``no two
processes ever execute their critical section simultaneously'' are,
shall I say, critical. Many approaches have been developed for
reasoning about programs using temporal logic. Most modern methods are
based on model checking (see \cite{r:clarke99}, for instance), while
other popular approaches are more proof-theoretic (see
\cite{r:schneider97}, for instance).

In the vast majority of cases, temporal logic is interpreted over
infinite state sequences. Those infinite sequences arise naturally,
for example, when modeling reactive systems, which are systems that
maintain a permanent interaction with their environment, and hence are
assumed to never terminate \cite{r:manna92}. Even when modeling
systems that may terminate, it is often acceptable to assume that the
final state of the system is simply infinitely repeated; this allows
infinite state sequences to be used. Intuitively, this approach works
as long as nothing of interest happens after the system has finished
executing. What happens, however, when one wants to reason about
explicitly finite state sequences? For instance, one may want to
reason about a sequence of states embedded in a larger structure,
where extending the sequence to an infinite sequence by repeating the
final state is not necessarily a reasonable step to take. This is
exactly what happens in \caret, where some of the temporal operators
are interpreted over the finite traces that make up procedure
invocations, all in the context of a complete program
execution.\footnote{Another context where this occurs is in process
logics \cite{r:pratt79}, which lets one reason about finite segments
of program executions within a larger and potentially infinite
execution. I hope to revisit this topic in an upcoming column.  Saake
and Lipeck \citeyear{r:saake88} and Havelund and Ro\c{s}u
\citeyear{r:havelund01} give additional motivation for considering
temporal reasoning over finite sequences.  Other uses of temporal
logic, for instance in descriptive complexity theory, often assume an
interpretation restricted to finite words
\cite{r:straubing94}.}

In order to characterize the properties of the \caret{} operators, we
need to understand the properties of temporal operators in the
presence of finite sequences.  Accordingly, my first goal is to make
clear the properties of temporal operators when interpreted over (1)
finite state sequences, (2) infinite state sequences, and (3) both
finite and infinite state sequences. To do this, I present a
particularly simple axiomatization of temporal logic that is sound and
complete over the class of finite and infinite state sequences. As
expected, a sound and complete axiomatization for the logic
interpreted over finite state sequences only can be derived by simply
adding an axiom that says ``there are no infinite state sequences'',
and a sound and complete axiomatization for the logic interpreted over
infinite state sequences only can be derived by simply adding an axiom
that says ``all state sequences are infinite''.  Interestingly, there
is a uniform elementary proof that covers all the cases. These results
can be found in various forms in the literature, albeit often
implicitly. 
The presentation I give is meant to emphasize the contribution of
exclusively finite and exclusively infinite traces to the
axiomatization of the temporal operators. The axiomatization will be
used as the basis of the sound and complete axiomatization for a
fragment of \caret{}.

\section*{Temporal Logic Over Infinite Sequences}

Let me first discuss finiteness and infinity
in the context of the simpler framework of propositional linear
temporal logic (LTL). The only temporal operators we consider are
future time operators, meaning that at a given state one can only
reason about the current and future states, and not past
states. Furthermore, LTL embodies a linear notion of time: from any
given state, there is a single sequence of states describing the
future.%
\footnote{This is in contrast to logics
interpreted over branching time, where a state can possibly have
multiple futures, and formulas can involve quantification over futures. 
See Emerson and Halpern \citeyear{r:emerson86} for
details on the relationship between linear and branching time temporal
logics.}

The language LTL is defined inductively by the following grammar,
where $p$ ranges over primitive propositions taken from a set
$\Phi_0$:
\[ \phi,\psi ::= p ~|~ \neg\phi ~|~ \phi\land\psi ~|~ \Next\phi ~|~
\phi\Until\psi. \]
Let $\phi\lor\psi$ stand for $\neg(\neg\phi\land\neg\psi)$, and
$\phi\rimp\psi$ stand for $\neg\phi\lor\psi$. Further, let
$\Sometimes\phi$ stand for $\truep\Until\phi$, and $\Always\phi$ stand
for $\neg\Sometimes\neg\phi$. Finally, define $\GNext\phi$ as the dual
of $\Next$, namely, $\neg\Next\neg\phi$. The operator $\Next$ is
sometimes called ``weak next''; $\Next\phi$ reads ``if there is a
next state, then $\phi$ holds there'', The operator $\GNext$ is
sometimes called ``strong next''; $\GNext\phi$ reads ``there is a
next state, and $\phi$ holds there''. The formula $\phi\Until\psi$
reads ``$\phi$ holds until $\psi$ is true'', while $\Sometimes\phi$
reads ``$\phi$ will eventually be true'' and $\Always\phi$ reads
``$\phi$ is and always will be true''. 

Temporal logic is interpreted over (linear) temporal structures.  A
\emph{temporal structure} is a tuple $M=(S,\sigma,\pi)$ where $S$ is a
set of states, $\sigma$ is a \emph{finite or infinite} sequence of
states in $S$, and $\pi$ is a valuation on the states, where $\pi(s)$
is the set of primitive propositions true at state $s$. Let $|\sigma|$
denote the length of $\sigma$, understood to be $\infty$ if $\sigma$
is infinite. Infinity is assumed to behave in the standard way with
respect to integers, for instance, $i < \infty$ for all integers
$i$. A temporal structure $M=(S,\sigma,\pi)$ is \emph{finite}
if $\sigma$ is finite, and \emph{infinite} otherwise. (Thus,
finiteness of a structure depends on the finiteness of the sequence,
not that of the state space.) If $\sigma=s_0s_1s_2\dots$, I will
sometimes use the notation $\sigma_i$ to refer to state $s_i$ in
$\sigma$. 

Let $\cM$ be the set of finite and infinite temporal structures. Let
$\cMinf$ be the class of infinite structures, and $\cMfin$ be the
class of finite structures. Satisfiability of a formula can be  defined in
a number of equivalent ways. If $M=(S,\sigma,\pi)$, where
$\sigma=s_0s_1\ldots$, possibly finite, define $(M,i)\sat\phi$,
meaning that formula $\phi$ is true in structure $M$ at position $i\in
\{0,\dots,|\sigma|\}$, inductively as follows:
\begin{itemize}
\item[] $(M,i)\sat p$ if $p\in\pi(s_i)$
\item[] $(M,i)\sat \neg\phi$ if $(M,i)\nsat\phi$
\item[] $(M,i)\sat \phi\land\psi$ if $(M,i)\sat\phi$ and
$(M,i)\sat\psi$
\item[] $(M,i)\sat \Next\phi$ if $i=|\sigma|$ or $(M,i+1)\sat\phi$
\item[] $(M,i)\sat \phi\Until\psi$ if $\exists j\in\{i,\dots,|\sigma|\}$ such that $(M,j)\sat\psi$ and $\forall
k\in\{i,\dots,j-1\}$, $(M,k)\sat\phi$.   
\end{itemize}
Observe that $\Next\phi$ is defined in such a way that if the
sequence is finite, $\Next\phi$ is true for all formulas $\phi$ at the
final state of the sequence. More drastically, $\Next\falsep$ is true
at a state if and only if it is the final state in the sequence. 
A formula $\phi$ is \emph{valid}, written $\sat\phi$, if
$(M,i)\sat\phi$ for all structures $M$ and positions $i$. 

The following axiomatization $\AX$ is well-known to be sound and
complete for temporal logic,\footnote{Recall that an axiomatization is
sound if every provable formula is valid, and complete if every valid
formula is provable.} as interpreted over infinite structures
\cite{r:gabbay80,r:fagin95,r:halpern04}: 
\begin{axiomlist}
\item[\axiom{Prop}.] All instances of propositional tautologies in
LTL.
\item[\axiom{MP}.] From $\phi$ and $\phi\rimp\psi$ infer $\psi$.
\item[\axiom{T1}.] $\Next\phi\land\Next(\phi\rimp\psi)\rimp\Next\psi$.
\item[\axiom{T2}.]
$\phi\Until\psi\riff\psi\lor(\phi\land\Next(\phi\Until\psi))$.
\item[\axiom{T3}.] $\Next(\neg\phi)\rimp\neg\Next\phi$.
\item[\axiom{RT1}.] From $\phi$ infer $\Next\phi$.
\item[\axiom{RT2}.] From $\phi'\rimp\neg\psi\land\Next\phi'$ infer
$\phi'\rimp\neg(\phi\Until\psi)$.
\end{axiomlist}

This axiomatization, by virtue of soundness and completeness,
intrinsically characterizes infinite structures. 
In fact, it is not hard to see that the axiomatization is not sound
for finite structures. More precisely, axioms \axiom{T2} and
\axiom{T3} are not valid in finite structures. To see this, let $p$
be a primitive proposition, and consider the structure
$M_1=(\{s\},s,\pi)$, that is, a finite structure with a single state
$s$, a sequence consisting of that single state $s$, and where
$\pi(s)=\{p\}$. It is easy to verify that 
\[ (M,0)\nsat \falsep\lor(p\land\Next(p\Until\falsep))\rimp p\Until\falsep,\]
which is an instance of \axiom{T2}, specifically, the $\Leftarrow$ implication
of \axiom{T2}, and
\[ (M,0)\nsat \Next\falsep\rimp\neg\Next\truep,\]
which is an instance of \axiom{T3}. Thus, in order to derive an axiomatization
that is sound and complete for a class structure including finite
ones, axioms \axiom{T2} and \axiom{T3} must somehow be weakened.
\section*{A General Axiomatization}

There is an axiomatization that is sound and complete for the
class of finite and infinite structures.  Let $\AXgen$ be the following
axiomatization, obtained from $\AX$ by replacing axioms \axiom{T2}
and \axiom{T3} by axioms
\axiom{T2'} and \axiom{T3'}:
\begin{axiomlist}
\item[\axiom{Prop}.] All instances of propositional tautologies in
LTL.
\item[\axiom{MP}.] From $\phi$ and $\phi\rimp\psi$ infer $\psi$.
\item[\axiom{T1}.] $\Next\phi\land\Next(\phi\rimp\psi)\rimp\Next\psi$.
\item[\axiom{T2'}.]
$\phi\Until\psi\riff\psi\lor(\phi\land\GNext(\phi\Until\psi))$.
\item[\axiom{T3'}.] $\Next\phi\riff(\Next\falsep\lor\GNext\phi)$.
\item[\axiom{RT1}.] From $\phi$ infer $\Next\phi$.
\item[\axiom{RT2}.] From $\phi'\rimp\neg\psi\land\Next\phi'$ infer
$\phi'\rimp\neg(\phi\Until\psi)$.
\end{axiomlist}
Axiom \axiom{T3'} captures the following intuition for $\Next\phi$:
either the next time step does not exist, or $\phi$ is true there. As
I have already argued, the fact that the next time step does not exist
is expressed by $\Next\falsep$.  The following variants of \axiom{T1}
are provable in $\AXgen$:
$\GNext\phi\land\Next(\phi\rimp\psi)\rimp\GNext\psi$, and
$\Next\phi\land\GNext(\phi\rimp\psi)\rimp\GNext\psi$. This
axiomatization is a simplification of the axiomatization of the future
fragment of the temporal logic of Lichtenstein, Pnueli, and Zuck
\citeyear{r:lichtenstein85}. Roughly speaking, the inference rule
\axiom{RT2} subsumes their axioms relating $\Next$ and $\Always$, using
the fact that $\Always$ is expressible using $\Until$. 

The following two axioms can be used to tailor the axiomatization to
the case where the structures are infinite, and the case where the
structures are finite. For infinite structures, an axiom is needed to
capture the fact that there is no final state: 
\begin{axiomlist}
\item[\axiom{Inf}.] $\neg\Next\falsep$.
\end{axiomlist}
To obtain an axiomatization for finite structures, an axiom is needed
to capture the fact that every finite structure has a final state:
\begin{axiomlist}
\item[\axiom{Fin}.] $\Sometimes\Next\falsep$.
\end{axiomlist}
Let $\AXinf$ be the axiomatization $\AXgen$ augmented with axiom
\axiom{Inf}, and let $\AXfin$ be the axiomatization $\AXgen$ augmented
with axiom \axiom{Fin}. These axiomatizations completely characterize
validity in the appropriate class of structures.  More precisely, the
following result holds. 
\begin{theorem}\label{t:main}
For formulas in the language LTL,
\begin{enumerate}
\item $\AXgen$ is a sound and complete axiomatization with respect
to $\cM$,
\item $\AXinf$ is a sound and complete axiomatization
with respect to $\cMinf$,
\item $\AXfin$ is a sound and complete axiomatization
with respect to $\cMfin$.
\end{enumerate}
\end{theorem}

The proof of this theorem is not difficult, and uses well-understood
technology. The only difficulty, in some sense, is coming up with the
proposed axiomatization. To illustrate where all the details are used,
let me spell out the details of the proof.
Soundness is straightforward to establish in all cases.  Completeness
is established by proving the following equivalent statement. Recall
that a formula $\phi$ is $\Ax$-consistent, for an axiomatization
$\Ax$, if $\neg\phi$ is not provable using the axioms and inference
rules of $\Ax$. Completeness is equivalent to the fact that
consistency implies satisfiability. Thus, it suffices to show that if
$\phi$ is consistent with respect to one of the particular
axiomatization, then it is satisfiable in a structure in the
corresponding class, that is, it is possible to construct an
appropriate structure such that $\phi$ is true in a state of the
structure.

The construction is essentially independent of the axiomatization
under consideration. Fix the formula $\phi$. The states of the model
will be constructed from an extension of the set of subformulas of
$\phi$. Let $\FL'(\phi)$ be the smallest set $S$ such that:
\begin{enumerate}
\item $\phi\in S$,
\item $\truep\Until\Next\falsep\in S$,
\item if $\neg\psi\in S$ then $\psi\in S$,
\item if $\psi_1\land\psi_2\in S$ then $\psi_1\in S$ and 
$\psi_2\in S$,
\item if $\Next\psi\in S$ then $\psi\in S$,
\item if $\Next\neg\psi\in S$ then $\Next\psi\in S$,
\item if $\psi_1\Until\psi_2\in S$ then $\psi_1\in S$,
$\psi_2 S$, and $\GNext(\psi_1\Until\psi_2) \in S$.
\end{enumerate}
Let $\FL(\phi)=\FL'(\phi)\cup\{\neg\psi\mid\psi\in\FL'(\phi)\}$. 
It is easy to check that for any $\phi$,
$\FL(\phi)$ is a finite set of formulas. Note that 
$\Next\falsep$ and $\neg\Next\falsep$ are always in $\FL(\phi)$. 

Let $\Ax$ range over $\AXgen$, $\AXinf$, and $\AXfin$. An $\Ax$-atom of
$\phi$ is a maximally $\Ax$-consistent subset of formulas in
$\FL(\phi)$.  It is easy to see that $\Ax$-atoms are finite. Let
$\At^{\Ax}(\phi)$ be the set of $\Ax$-atoms of $\phi$; we use
$V,W,\dots$ to denote $\Ax$-atoms.  Associate with every $\Ax$-atom $V$
a formula $\widehat{V}$, the conjunction of all the 
formulas in $V$, that is, $\widehat{V}=\bigvee_{\psi\in V}\psi$. 
It is straightforward to check that for every formula
$\psi\in\FL(\phi)$ and every $\Ax$-atom $V$ of $\phi$, either $\psi$
or $\neg\psi$ is in $V$. (If not, then $V$ is not maximally
$\Ax$-consistent.) Using axiom $\axiom{Prop}$, it is easy to show that
any formula $\psi\in\FL(\phi)$ is provably equivalent to the
disjunction $\bigvee_{\{V\in\At^{\Ax}\mid\psi\in V\}}\widehat{V}$, and
$\truep$ is provably equivalent to the disjunction
$\bigvee_{V\in\At^{\Ax}}\widehat{V}$.

\newcommand{\lra}[1]{\stackrel{#1}{\longrightarrow}}
\newcommand{\lr}{\longrightarrow}

For $\Ax$-atoms $V$ and $W$, define $V\lra{\Ax} W$ if
$\widehat{V}\land\GNext \widehat{W}$ is $\Ax$-consistent. Let
$V\mathord{\lra{\Ax}}$ be the set $\{W\mid
V\lra{\Ax}W\}$. A \emph{chain} of $\Ax$-atoms is a finite or
infinite sequence $V_0,V_1,\dots$ of $\Ax$-atoms with the property
that $V_i\lra{\Ax}V_{i+1}$, for all $i$. A chain $V_0,V_1,\dots$ of $\Ax$-atoms
is \emph{acceptable} if for all $i$, whenever
$\psi_1\Until\psi_2\in V_i$, then there exists
$j\ge i$ such that $\psi_2\in V_j$ and $\psi_1\in
V_i,\dots,V_{j-1}$.  The following lemma isolates all
the properties needed to prove the completeness results.
\begin{lemma}\label{l:1}
\begin{enumerate}
\item 
For all $\Next\psi\in\FL(\phi)$ and $\Ax$-atoms $V$, $\Next\psi\in 
V$ if and
only if for all $W\in V\mathord{\lra{\Ax}}$, $\psi\in W$
\item 
For all $\GNext\psi\in\FL(\phi)$ and $\Ax$-atoms $V$,
$\GNext\psi\in V$ if and only if there exists
$W\in V\mathord{\lra{\Ax}}$ such that $\psi\in W$.  
\item 
For all $\psi_1\Until\psi_2\in\FL(\phi)$ and $\Ax$-atoms $V_0$,
$\psi_1\Until\psi_2\in V_0$ if and only if there exists a finite chain 
$V_0,V_1,\dots,V_k$ such that $\psi_1\in
V_0,\dots,V_{k-1}$ and $\psi_2\in V_k$.  
\item For all $\Ax$-atoms $V$, $\Next\falsep\in V$ if and only if
$V\mathord{\lra{\Ax}}=\emptyset$. 
\item For all $\AXfin$-atoms $V_0$, there exists a finite chain
$V_0,\dots,V_k$ such that $\Next\falsep\in V_k$. 
\item Every finite chain of $\AXgen$-atoms is extensible to an
acceptable chain (finite or infinite). 
\item Every finite chain of $\AXinf$-atoms is extensible to an
infinite acceptable chain.
\item Every finite chain of $\AXfin$-atoms is extensible to a
finite acceptable chain. 
\end{enumerate}
\end{lemma}
\begin{proof}
The proof
technique is adapted from that of Halpern, van der Meyden, and Vardi
\citeyear{r:halpern04}.

(a) Assume that $\Next\psi\in V$, and let $W\in V\mathord{\lra{\Ax}}$. By way
of contradiction, assume that $\psi\not\in W$. Then, 
$\neg\psi\in W$, that is, $\vdash \widehat{W}\rimp\neg\psi$. By
\axiom{Prop} and \axiom{RT1}, $\vdash\Next(\psi\rimp\neg\widehat{W})$. By
assumption, $\Next\psi\in V$,
that is, $\vdash \widehat{V}\rimp\Next\psi$. By \axiom{MP} and \axiom{T1}, $\vdash
\widehat{V}\rimp\Next\neg\widehat{W}$. But $V\lra{\Ax} W$ means that
$\widehat{V}\land\GNext\widehat{W}$ is consistent, so that $\not\vdash
\widehat{V}\rimp\Next\neg\widehat{W}$, a contradiction. So $\psi\in W$. 

Conversely, assume that for all $W\in V\mathord{\lra{\Ax}}$, $\psi\in W$. By
way of contradiction, assume that $\Next\psi\not\in V$, so that
$\neg\Next\psi\in V$, and thus $\vdash \widehat{V}\rimp\neg\Next\psi$. For
any $W$ such that $\psi\not\in W$, it must be the case that $W\not\in
V\mathord{\lra{\Ax}}$, and thus $\widehat{V}\land\GNext\widehat{W}$ is
inconsistent. Thus, $\widehat{V}\land\GNext\widehat{W}$ is inconsistent for all
$W$ such that $\psi\not\in W$, and $\lor_{\{W\mid \psi\not\in
W\}}(\widehat{V}\land\GNext\widehat{W})$, is inconsistent, that is,
$\widehat{V}\land\GNext\neg\psi$ is inconsistent, and $\vdash
\widehat{V}\rimp\neg\GNext\neg\psi$, or $\vdash\widehat{V}\rimp\Next\psi$. By
assumption, $\vdash\widehat{V}\rimp\neg\Next\psi$, so that
$\vdash\widehat{V}\rimp\falsep$, that is, $\vdash\neg\widehat{V}$, which
contradicts the fact that $V$ is a consistent set of formulas. Thus,
$\Next\psi\in V$, as desired.

(b) Assume that $\GNext\psi\in V$. If
$\GNext\psi\in V$, then $\Next\psi\in V$, and $\Next\psi\in\FL(\phi)$
by closure rule (4). Hence, by part (a), all $W\in
V\mathord{\lra{\Ax}}$ are such that $\psi\in W$. It suffices to show then there is a $W$ such that
$V\lra{\Ax} W$. Assume not. Then $\lor_{\{W\mid\truep\in W\}}(\widehat{V}\land\GNext
\widehat{W})$ is inconsistent, and hence $\vdash\neg(\widehat{V}\land\GNext\truep)$, and
$\vdash \widehat{V}\rimp\Next\falsep$. Because $\GNext\psi\in V$, then
$\vdash \widehat{V}\rimp\neg\Next\falsep$. So $\vdash
\widehat{V}\rimp\falsep$, that is, $\vdash\neg \widehat{V}$, contradicting $V$ being
consistent. So there must be a $W\in V\mathord{\lra{\Ax}}$. 

Conversely, assume that there exists $W\in V\mathord{\lra{\Ax}}$ and $\psi\in
W$. Since $\widehat{V}\land\GNext \widehat{W}$ is consistent, so is
$\lor_{\{W\mid \psi\in W\}}(\widehat{V}\land\GNext \widehat{W})$, and
$\widehat{V}\land\GNext\psi$. In other words, $\not\vdash
\widehat{V}\land\neg\GNext\psi$. Assume by way of contradiction that
$\GNext\psi\not\in V$, so that $\neg\GNext\psi\in V$. Then $\vdash
\widehat{V}\rimp \neg\GNext\psi$, a contradiction. Therefore,
$\GNext\psi\in V$.

(c) 
Assume that $\psi_1\Until\psi_2\in V_0$. Suppose by way of
contradiction that no suitable chain exists. Let $T$ be the
smallest set $S$ of $\Ax$-atoms of $\phi$ such that $V_0\in S$, and
if $W\in V\mathord{\lra{\Ax}}$ (for some $V$ in $S$) and $\widehat{W}\land\psi_1$,
then $W\in S$. If $T$ is a set of $\Ax$-atoms, let
$\widehat{T}=\bigvee_{W\in T}\widehat{W}$. Clearly, 
$\neg\psi_2\in W$ for all $W$ in $T$, and thus, $\vdash \widehat{T}\rimp\neg\psi_2$. Moreover, for every $V$ in $T$ and $W\in
V\mathord{\lra{\Ax}}$, either $W\in T$, or $\neg\psi_1\in W$ and
$\neg\psi_2\in W$. This yields $\vdash
\widehat{T}\rimp\Next(\widehat{T}\lor(\neg\psi_1\land\neg\psi_2))$. It
follows easily from 
\axiom{T1}, \axiom{T2'}, \axiom{RT1}, \axiom{RT2} that $\vdash
\widehat{T}\rimp\neg(\psi_1\Until\psi_2)$. In particular, $\vdash 
\widehat{V_0}\rimp\neg(\psi_1\Until\psi_2)$, contradicting
$\psi_1\Until\psi_2\in V_0$. 

Conversely, by induction on $k$, if there exists $V_1\in
V_0\mathord{\lra{\Ax}},\dots,V_k\in V_{k-1}\mathord{\lra{\Ax}}$, $\psi_1\in V_i$ for
$i\in\{0,\dots,k-1\}$, and $\psi_2\in V_k$, then
$\psi_1\Until\psi_2\in V_0$. If $k=0$, the result follows immediately
by an application of \axiom{T2'} and \axiom{T3'}. For a general $k$,
assume by way of contradiction that $\psi_1\Until\psi_2\not\in V_0$
(so that $\neg\psi_1\Until\psi_2\in V_0$), and 
consider the subchain $V_1,\dots,V_k$, such that $V_2\in
V_1\mathord{\lra{\Ax}},\dots,V_k\in V_{k-1}\mathord{\lra{\Ax}}$, $\psi_1\in
V_1,\dots,V_{k-1}$, and $\psi_2\in V_k$. By the induction hypothesis,
$\psi_1\Until\psi_2\in V_1$, that is, $\vdash
\widehat{V_1}\rimp\psi_1\Until\psi_2$. Since $V_1\in V_0\mathord{\lra{\Ax}}$, 
$\widehat{V_0}\land\GNext\widehat{V_1}$ consistent, and by an
application of \axiom{RT1} and a $\GNext$-variant of \axiom{T1},
$\widehat{V_0}\land\GNext\psi_1\Until\psi_2$ is consistent. Since
$\psi_1\in V_0$, $\vdash \widehat{V_0}\rimp\psi_1$, and thus
$\widehat{V_0}\land\psi_1\land\GNext\psi_1\Until\psi_2$ is also
consistent. By \axiom{T2'}, $\widehat{V_0}\land\psi_1\Until\psi_2$ is
consistent, that is,
$\not\vdash\widehat{V_0}\rimp\neg\psi_1\Until\psi_2$, contradicting the
assumption that $\neg\psi_1\Until\psi_2\in V_0$. Thus,
$\psi_1\Until\psi_2$ must be in $V_0$, as desired.  

(d) Assume that $\vdash\widehat{V}\rimp\Next\falsep$. By way of
contradiction, assume there is a $W\in V\mathord{\lra{\Ax}}$. By
\axiom{Prop}, $\vdash \falsep\rimp\neg\widehat{W}$, and by
\axiom{RT1}, $\vdash\Next(\falsep\rimp\neg\widehat{W})$. By propositional
reasoning and \axiom{T1}, $\vdash\widehat{V}\rimp\Next\neg\widehat{W}$,
which is equivalent to $\vdash\widehat{V}\rimp\neg\GNext\widehat{W}$, that is,
$\vdash\neg(\widehat{V}\land\GNext\widehat{W})$, contradicting the assumption
that $W\in V\mathord{\lra{\Ax}}$, that is, that $\widehat{V}\land\GNext\widehat{W}$ is
consistent. 

Conversely, assume that there is no $W\in
V\mathord{\lra{\Ax}}$. Therefore, for all $W$, $\widehat{V}\land\GNext\widehat{W}$ is
inconsistent, and thus, $\lor_{W}(\widehat{V}\land\GNext\widehat{W})$ is
inconsistent. By propositional reasoning, 
$\widehat{V}\land\GNext\lor_W\widehat{W}$, and thus $\widehat{V}\land\GNext\truep$, is 
inconsistent. By propositional reasoning and definition of $\GNext$,
this simply means that $\vdash\widehat{V}\rimp\Next\falsep$. 

(e) Let $V_0$ be an $\AXfin$-atom. Suppose by way of
contradiction that no suitable chain exists. Let $T$ be the
smallest set $S$ of $\AXfin$-atoms of $\phi$ such that $V_0\in S$, and
if $W\in V\mathord{\lra{\AXfin}}$ (for some $V$ in $S$) then $W\in S$. Clearly,
$\neg\Next\falsep\in W$ for all $W$ in $T$ (otherwise, it could be used to 
construct a finite chain assumed not to exist), 
and thus, $\vdash \widehat{T}\rimp\neg\Next\falsep$. Moreover, for every $V$ in $T$ and $W\in
V\mathord{\lra{\AXfin}}$, $W\in T$. Therefore, it is possible to derive
$\vdash \widehat{T}\rimp\Next\widehat{T}$, which implies that $\vdash
\widehat{T}\rimp\Next(\widehat{T}\lor(\neg\truep\land\neg\Next\falsep))$.
It follows easily from 
\axiom{T1}, \axiom{T2'}, \axiom{RT1}, \axiom{RT2} that $\vdash
\widehat{T}\rimp\neg(\truep\Until\Next\falsep)$. In particular,
$\vdash \widehat{V_0}\rimp\neg(\truep\Until\Next\falsep)$,
contradicting $\truep\Until\Next\falsep\in V_0$, by virtue of
axiom \axiom{Fin}.  

(f) Let $V_0,\dots,V_n$ be a finite chain of $\AXgen$-atoms. Consider a
formula $\psi_1\Until\psi_2\in V_0$. It follows, from \axiom{T2'} and
parts (a) and (b), either that $\psi_2\in V_j$ for some
$j\in\{0,\dots,n\}$ and $\psi_1\in V_l$ for $l\in\{0,\dots,j-1\}$, or
that $\psi_1\in V_j$ for all $j\in\{0,\dots,n\}$, and
$\psi_1\Until\psi_2\in V_n$. In the latter case, by
part (c), there exists a chain $V_n,\dots,V_{n'}$ such that
$\psi_1\in V_k$ for $k\in\{n,\dots,n'-1\}$ and $\psi_2\in
V_{n'}$. This gives a finite extension of the original chain that
satisfies the obligation of acceptability for $\psi_1\Until\psi_2$ at
$V_0$. Applying this argument to the remaining $\Until$-formulas in
$V_0$ produces a finite chain that satisfies all the
obligations at $V_0$. Apply the same procedure to $V_1$, and so
on. In the limit, this produces an acceptable chain extending the
original chain. This chain can be either finite, or infinite.  

(g) Let $V_0,\dots,V_n$ be a finite chain of $\AXinf$-atoms. Just as 
in part (f), it is possible to construct an acceptable chain extending this
chain that satisfies all the obligations of the $\Until$-formulas. If
this process results in a finite acceptable chain $V_0,\dots,V_{n'}$,
this chain can be extended to an infinite acceptable chain as
follows. Given the final state $V_{n'}$ of the chain, there exists a
state $V_{n'+1}\in V_{n'}\mathord{\lra{\AXinf}}$. Otherwise, by
part (d), $\vdash_{\AXinf}\widehat{V_{n'}}\rimp\Next\falsep$.
However, by \axiom{Inf}, $\vdash_{\AXinf}\neg\Next\falsep$, and thus by 
\axiom{MP}, $\vdash_{\AXinf}\neg\widehat{V_{n'}}$, contradicting the fact 
that $V_{n'}$ is $\AXinf$-consistent. Thus, there must exist
$V_{n'+1}\in V_{n'}\mathord{\lra{\AXinf}}$. Let $V_0,\dots,V_{n'+1}$
be the new chain formed in this way. This chain can be once again
extended to an acceptable chain, by ensuring that all the obligations
of the $\Until$-formulas are satisfied. In the limit, this new
procedure produces an infinite acceptable chain.

(h) Let $V_0,\dots,V_n$ be a finite chain of $\AXfin$-atoms,
$V_0,\dots,V_k$. By part (e), there exists a finite chain
$V_k,\dots,V_n$ such that $\Next\falsep\in V_n$. By part (d), this
means that there $V_n\mathord{\lra{\AXfin}}=\emptyset$. It remains to show that
the chain is acceptable, that is, for every $\psi_1\Until\psi_2$ in
$V_0,\dots,V_n$, the obligations are met. Let
$\psi_1\Until\psi_2\in V_i$. Just as in part (f), it follows, from
\axiom{T2'} and parts (a) and (b), either that $\psi_2\in V_j$ for
some $i\le j\le n$ and $\psi_1\in V_l$ for $i\le l < j$, or
that $\psi_1\in V_j$ for all $i\le j \le n$, and both
$\neg\psi_2\in V_n$ and  $\psi_1\Until\psi_2\in V_n$. In the former
case, the obligations for $\psi_1\Until\psi_2$ are met. The latter
case cannot arise. Indeed, if $\psi_1\Until\psi_2\in V_n$,
then $\vdash \widehat{V_n}\rimp\psi_1\Until\psi_2$, so that $\vdash
\widehat{V_n}\rimp\psi_2\lor(\psi_1\land\GNext(\psi_1\Until\psi_2))$.
Since $\neg\psi_2\in V_n$, $\vdash \widehat{V_n}\rimp\neg\psi_2$, so that
$\vdash
\widehat{V_n}\rimp\psi_1\land\GNext(\psi_1\Until\psi_2)$. Therefore, 
$\GNext(\psi_1\Until\psi_2)$ must be in $V_n$. By part (b), there must
exist $W\in V_n\mathord{\lra{\AXfin}}$ with $\psi_1\Until\psi_2\in W$, which
contradicts that fact that $V_n\mathord{\lra{\AXfin}}=\emptyset$. 
\end{proof}

The completeness results of Theorem~\ref{t:main} follow easily from
Lemma~\ref{l:1}. Consider the axiomatization $\AXgen$. Assume that
$\phi$ is $\AXgen$-consistent. Since $\phi\in\FL(\phi)$, $\phi\in
V^\phi$ for some $\AXgen$-atom $V^\phi$ of $\phi$.  Construct the
structure $M=(S,\sigma,\pi)$ by taking the set of states $S$ to be the
set $\At^{\AXgen}(\phi)$ of $\AXgen$-atoms of $\phi$. Define the
interpretation $\pi$ by $\pi(V)=\{p\mid p\in V\}$. All that remains
now is to extract a sequence $\sigma$ in $S$ that satisfies
$\phi$. By Lemma~\ref{l:1}(f), $V^\phi$, a one-element finite chain of
$\AXgen$-atoms, is extensible to an acceptable chain $\sigma=V_0
V_1\dots$. It is easy to check, by induction on the structure of
$\phi$, that $(M,i)\sat\phi$ if and only if $\phi\in V_i$. Since
$\phi\in V^\phi=V_0$, then $(M,0)\sat\phi$. A similar argument
holds for $\AXinf$ and $\AXfin$, invoking Lemma~\ref{l:1}(g) and
Lemma~\ref{l:1}(h), respectively, to construct an acceptable chain
$\sigma$.

\section*{The Linear Temporal Logic of Calls and Returns}

While the above discussion is still fresh, let me now talk about the
\caret{} logic. \caret{} was designed for reasoning about programs, in
the form of state sequences, each sequence corresponding to an
execution of the program. It was especially designed for
reasoning about \emph{nonregular} properties of programs. The
classical example of such a property is the correctness of procedures
with respect to pre and post conditions, that is, verifying that if
$\phi$ holds before every call to a procedure, then $\psi$ holds after
the procedure returns. The nonregularity of this property is due to
the fact that finding the state where the procedure returns requires
matching the number of calls and returns within the body of the
procedure. \caret{} provides operators for doing just that. While
frameworks for verifying procedure with respect to pre and post
conditions go back to the seminal work of Hoare \citeyear{r:hoare69},
the main contribution of
\caret{} is a decidable model-checking procedure for programs
expressed as recursive state machines (equivalently, pushdown systems)
\cite{r:alur01,r:benedikt01}. 
To achieve this, \caret{} assumes that every state is tagged,
indicating whether it is a call state (meaning it is a state that
performs a procedure call), a return state (meaning it is a state that
corresponds to having returned from a procedure), or an internal state
(everything else). I will not discuss the model-checking algorithm
here, but instead examine the properties of the new operators
that \caret{} introduces. 

The language \caret{} of linear propositional temporal logic with
calls and returns is defined inductively by the following grammar,
where $p$ ranges over primitive propositions taken from a set
$\Phi_0$, which includes $\call$,$\ret$, and $\inter$:\footnote{In
fact, this is just a fragment of \caret{}. The full logic includes
past-time temporal operators that walk back the call chain of a
procedure. I believe that the development in this section extends in a
straightforward way to the full language, but I have not checked the
details.}
\[ \phi,\psi ::= p ~|~ \neg\phi ~|~ \phi\land\psi ~|~ \Next\phi ~|~
\phi\Until\psi ~|~ \Next^a\phi ~|~ \phi\Until^a\psi. \]
As before, define the usual abbreviations. Let $\phi\lor\psi$ stand
for $\neg(\neg\phi\land\neg\psi)$, and $\phi\rimp\psi$ stand for
$\neg\phi\lor\psi$. Define, as in LTL, $\Sometimes\phi$ to stand for
$\truep\Until\phi$, $\Always\phi$ to stand for
$\neg\Sometimes\neg\phi$, and $\GNext\phi$ to stand for
$\neg\Next\neg\phi$. Define $\Sometimes^a\phi$, $\Always^a\phi$, and
$\GNext^a\phi$ in a similar way. 

The $\Next$ and $\Until$ operators, the global-time operators, are the
standard operators from LTL, interpreted over whole sequences of
states.\footnote{Alur, Etessami, and Madhusudan 
use $\Next^g\phi$ and $\phi\Until^g\psi$ for the global-time
operators.} Thus, $\Next\phi$ means that $\phi$ holds at the next
state whether or not the next state is a state in an invoked
procedure, or the next state follows from returning from a
procedure. The $\Next^a$ and $\Until^a$ operators, the abstract-time
operators, do not consider all states in the sequence, but only the
states in the current procedure context. Thus, $\Next^a\phi$ means
that $\phi$ holds at the abstract next state of the procedure---if the
current state is a procedure call, then the abstract next state is in
fact the matching return state; if the current state is the last state
of a procedure invocation, there is no abstract next state; similarly,
if the current state is a procedure call that never returns (say, it
enters an infinite loop), there is no abstract next
state. Correspondingly, $\phi\Until^a\psi$ means that the abstract
path from the current state (i.e., the path formed by successive abstract
successors) satisfies $\phi\Until\psi$. 

To formalize these intuitions, \caret{} is interpreted over structured
(linear) temporal structures.  An \emph{structured temporal structure}
is a tuple $M=(S,\sigma,\pi)$ where $S$ is a set of states, $\sigma$
is an infinite sequence of structured states in
$S\times\{\call,\ret,\inter\}$, and $\pi$ is a valuation on the
states, where $\pi(s)$ is the set of primitive propositions true at
state $s$. For a sequence $\sigma$, define an abstract successor
function $\mathit{succ}^a_\sigma$ giving, for every index $i$ into
$\sigma$, the index of the next abstract state for
$\sigma_i$. Formally, the abstract successor is defined as
follows. First, for a sequence of structured states $\sigma$, define
the partial map $R_\sigma(i)$, which maps any $i$ to the first
unmatched return after $i$, that is, the first return that does not
correspond to a procedure call performed after $i$: $R_\sigma(i)=j$,
where $j$ is the smallest $j'$ such that $j'>i$, $\sigma_{j'}$ is a
return state, and the number of calls and returns in
$\sigma_{i+1},\dots,\sigma_{j'-1}$ are equal; $R_\sigma(i)=\bot$ if
there is no such $j'$. (Intuitively, $\bot$ represents the value
``undefined''.) The abstract successor functions can now be defined:
\[ \mathit{succ}^a_\sigma(i) \triangleq 
 \begin{cases}
   R_\sigma(i) & \text{if $\sigma_i=(-,\call)$}\\
   \bot & \text{if $\sigma_i\ne(-,\call)$ and $\sigma_{i+1}=(-,\ret)$}\\
   i+1 & \text{otherwise.}
 \end{cases}
\]

Let $\cM^\mathit{cr}$ be the set of structured temporal
structures. Satisfiability of a formula is defined as follows. If
$M=(S,\sigma,\pi)$, where $\sigma=\<(s_0,t_0),(s_1,t_1),\ldots\>$, 
define $(M,i)\sat\phi$, 
meaning that formula $\phi$ is true in structure $M$ at position $i\ge
0$ inductively as follows:
\begin{itemize}
\item[] $(M,i)\sat p$ if $p\in\pi(s_i)$ or $p=t_i$
\item[] $(M,i)\sat \neg\phi$ if $(M,i)\nsat\phi$
\item[] $(M,i)\sat \phi\land\psi$ if $(M,i)\sat\phi$ and
$(M,i)\sat\psi$
\item[] $(M,i)\sat \Next\phi$ if $(M,i+1)\sat\phi$
\item[] $(M,i)\sat \phi\Until\psi$ if  $\exists j\ge i$
such that $(M,j)\sat\psi$ and $\forall k\in\{i,\dots,j-1\}$, $(M,k)\sat\phi$
\item[] $(M,i)\sat \Next^a\phi$ if $\mathit{succ}^a_\sigma(i)=\bot$ or
$(M,\mathit{succ}^a_\sigma(i))\sat\phi$
\item[] $(M,i)\sat \phi\Until^a\psi$ if 
$\exists i_0,i_1,\dots,i_k$ (with $i_0=i$)
such that $\mathit{succ}^a_\sigma(i_j)=i_{j+1}\ne\bot$ (for $j=0,\dots,k-1$),
$(M,i_k)\sat\psi$, and
$(M,i_j)\sat\phi$ (for $j=0,\dots,k-1$). 
\end{itemize}
(The semantics above uses a ``weak'' semantics for $\Next^a\phi$,
while the original description of \caret{} uses a ``strong''
semantics. In other words, the interpretation of $\Next^a\phi$ in the
original \caret{} is the same as that of $\GNext^a\phi$ here. I made
this choice for consistency with the usual reading of $\Next$ as a
weak next and to reuse the development of last section. Clearly, there
is no loss of expressiveness from this change.) 

What about an axiomatization for this logic? The following axioms
account for the fact that the $\Next/\Until$ fragment of \caret{} is
essentially LTL interpreted over infinite sequences.
\begin{axiomlist}
\item[\axiom{Prop}.] All instances of propositional tautologies in
\caret.
\item[\axiom{MP}.] From $\phi$ and $\phi\rimp\psi$ infer $\psi$.
\item[\axiom{G1}.] $\Next\phi\land\Next(\phi\rimp\psi)\rimp\Next\psi$.
\item[\axiom{G2}.]
$\phi\Until\psi\riff\psi\lor(\phi\land\GNext(\phi\Until\psi))$.
\item[\axiom{G3}.] $\Next\phi\riff(\Next\falsep\lor\GNext\phi)$.
\item[\axiom{G4}.] $\neg\Next\falsep$.
\item[\axiom{RG1}.] From $\phi$ infer $\Next\phi$.
\item[\axiom{RG2}.] From $\phi'\rimp\neg\psi\land\Next\phi'$ infer
$\phi'\rimp\neg(\phi\Until\psi)$.
\end{axiomlist}

The operators $\Next^a$ and $\Until^a$ behave like the standard
temporal operators, except they are interpreted over possibly finite
sequences.
\begin{axiomlist}
\item[\axiom{A1}.] $\Next^a\phi\land\Next^a(\phi\rimp\psi)\rimp\Next^a\psi$.
\item[\axiom{A2}.]
$\phi\Until^a\psi\riff\psi\lor(\phi\land\GNext^a(\phi\Until^a\psi))$.
\item[\axiom{A3}.] $\Next^a\phi\riff(\Next^a\falsep\lor\GNext^a\phi)$.
\item[\axiom{RA1}.] From $\phi$ infer $\Next^a\phi$.
\item[\axiom{RA2}.] From $\phi'\rimp\neg\psi\land\Next^a\phi'$ infer
$\phi'\rimp\neg(\phi\Until^a\psi)$.
\end{axiomlist}

The remaining axioms capture the relationship between the kind of
states (call states, return states, internal states), and the
behavior of the various next-time operators. Roughly, this amounts to
capturing the properties of the $\mathit{succ}^a_\sigma$ function,
when it is defined, and when it is not. The following axiom says that
there is exactly one of $\call,\ret,\inter$ that holds at any state.
\begin{axiomlist}
\item[\axiom{C1}.]
$(\call\land\neg\ret\land\neg\inter)\lor(\neg\call\land\ret\land\neg\inter)\lor(\neg\call\land\neg\ret\land\inter)$.
\end{axiomlist}
If the current state is not a call state, then the properties of the
abstract next state operator depend on whether the next global state
is a return state.
\begin{axiomlist}
\item[\axiom{C2}.]
$\neg\call\land\Next(\neg\ret)\rimp(\Next\phi\riff\GNext^a\phi)$.
\item[\axiom{C3}.]
$\neg\call\land\Next(\ret)\rimp\Next^a\falsep$.
\item[\axiom{C4}.]
$\GNext^a\phi\rimp\Sometimes\phi$.
\end{axiomlist}
Already, it is possible to derive from these axioms that if a state is
not a call state and there is no abstract next state, then the global
next state must be a return state; in other words, the only case where
there is no abstract next state (unless a procedure call is performed)
is at the end of a procedure invocation. Here is a formal derivation of
$\neg\call\land\Next^a\falsep\rimp\Next\ret$:
\begin{alignat*}{3}
 & 1. &\quad & \vdash \neg\call\land\Next\neg\ret\rimp(\Next\truep\riff\neg\Next^a\falsep) &\quad& (\axiom{C2})\\
 & 2. && \vdash \neg\call\rimp(\Next\neg\ret\rimp(\Next\truep\rimp\neg\Next^a\falsep)) && (1,\axiom{Taut},\axiom{MP})\\
 & 3. && \vdash (\Next\neg\ret\rimp(\Next\truep\rimp\neg\Next^a\falsep))\rimp(\Next\truep\rimp(\Next^a\falsep\rimp\neg\Next\neg\ret)) && (\axiom{Taut})\\
 & 4. && \vdash \neg\call\rimp(\Next\truep\rimp(\Next^a\falsep\rimp\neg\Next\neg\ret)) && (2,3,\axiom{MP})\\
 & 5. && \vdash \Next\truep\rimp(\neg\call\rimp(\Next^a\falsep\rimp\neg\Next\neg\ret)) && (4,\axiom{Taut},\axiom{MP})\\
 & 6. && \vdash \Next\truep && (\axiom{Taut},\axiom{RG1})\\
 & 7. && \vdash \neg\call\rimp(\Next^a\falsep\rimp\neg\Next\neg\ret) && (5,6,\axiom{MP})\\
 & 8. && \vdash \neg\call\land\Next^a\falsep\rimp\neg\Next\neg\ret && (7,\axiom{Taut},\axiom{MP})\\
 & 9. && \vdash \Next\ret\riff(\Next\falsep\lor\neg\Next\neg\ret) && (\axiom{G3})\\
 & 10. && \vdash (\Next\ret\riff(\Next\falsep\lor\neg\Next\neg\ret))\rimp (\neg\Next\falsep\rimp(\neg\Next\neg\ret\rimp\Next\ret)) && (\axiom{Taut})\\
 & 11. && \vdash \neg\Next\falsep\rimp(\neg\Next\neg\ret\rimp\Next\ret) && (10,\axiom{MP})\\
 & 12. && \vdash \neg\Next\falsep && (\axiom{G4})\\
 & 13. && \vdash \neg\Next\neg\ret\rimp\Next\ret && (11,12,\axiom{MP})\\
 & 14. && \vdash \neg\call\land\Next^a\falsep\rimp\Next\ret && (8,13,\axiom{MP}).
\end{alignat*}

If the current position is a call, then the abstract successor exists
or not, depending on whether or not there is a balanced number of
calls and returns before the return matching the call. This turns out
to be painful to capture. Intuitively, the logic cannot count---there
is no way to say (directly) that ``there are exactly the same number
of call states as there are return states before the matching
return''. The best one can do is basically enumerate all
possibilities. Define the class of formulas
$\mathit{CR}^c_{m,n}(\phi)$, one for every $c,m,n\ge 0$ such that
$c+m\ge n$. Intuitively, $\mathit{CR}^c_{m,n}(\phi)$ says that between
the current state and the first state where $\phi$ holds, there are
exactly $m$ call states and $n$ return states, and moreover there are
never more then $c$ return states more than the number of call states.
\[
\mathit{CR}^c_{m,n}(\phi)\triangleq
\begin{cases}
 \inter\Until\phi & \text{if $m=0, n=0$}\\
 \inter\Until(\call\land\Next\mathit{CR}^{c+1}_{m-1,n} & \text{if $m>0, n=0$}\\
 \inter\Until(\ret\land\Next\mathit{CR}^{c-1}_{m,n-1} & \text{if $m=0, n>0$}\\
 \inter\Until(\call\land\Next\mathit{CR}^{c+1}_{m-1,n} & \text{if $m>0, n>0, c=0$}\\
 \begin{prog}
   \inter\Until(\call\land\Next\mathit{CR}^{c+1}_{m-1,n}(\phi))\\
   \quad\lor\inter\Until(\ret\land\Next\mathit{CR}^{c-1}_{m,n-1}(\phi))
 \end{prog} & \text{if $m>0, n>0, c>0$.}
\end{cases}
\]
With this, it is possible to define a countable number of axioms that
essentially say that if the current state is a call state, and the
number of calls and returns following the current state match before
there is a return where $\phi$ holds, then there is an abstract
next state and $\phi$ holds there.
\begin{axiomlist}
\item[\axiom{C5}.]
$\call\land\Next\mathit{CR}^0_{n,n}(\ret\land\phi)\rimp\GNext^a\phi$
(for $n\ge 0$).
\end{axiomlist}
Of course, \axiom{C5} is a family of axioms, one for each $n\ge 0$.

Similarly, if the number of calls exceeds the number of returns after
a call state, then there cannot be an abstract next state.
\begin{axiomlist}
\item[\axiom{C6}.]
$\call\land\Next\mathit{CR}^0_{m,n}(\Always\neg\ret)\rimp\Next^a\falsep$
(for $m>n\ge 0$). 
\end{axiomlist}
Again, \axiom{C6} is a family of axioms, one for each $m>n\ge 0$.

\begin{theorem}\label{t:caret}
$\AXcr$ is a sound and complete axiomatization for \caret{} with respect
to structured temporal structures. 
\end{theorem}

Soundness is straightforward. The strategy for proving completeness
is, unsurprisingly, analogous to that of the proof of
Theorem~\ref{t:caret}: assuming $\phi$ is consistent, completeness
requires showing that $\phi$ is satisfiable; to construct a model of
$\phi$, take an atom of $\phi$ containing $\phi$, and extend it to an
acceptable infinite sequence of atoms. I leave it to the reader to
generalize $\FL'(\phi)$ and $\FL(\phi)$ in the right way, by adding
clauses for $\Next^a$ and $\Until^a$ mimicking those for $\Next$ and
$\Until$. Atoms of $\phi$ have the same definition, maximally
consistent subsets of formulas in $\FL(\phi)$, as does the formula
$\widehat{V}$ for any given atom $V$. (For atoms, as for the other
notions, there is no need for an axiomatization qualification,
as there is a single axiomatization to consider.)  The relation $V\lr
W$ between atom is again defined to hold if
$\widehat{V}\land\GNext\widehat{W}$ is consistent. A \emph{chain} of atoms is a
finite or infinite sequence $V_0,V_1,\dots$ of atoms with the property
that $V_i\lr V_{i+1}$, for all $i$. It remains to show how to
extend a finite chain of atoms into a suitably defined acceptable
chain. This is where things vary from the proof of
Theorem~\ref{t:main}, to account for the structured states.

First, given a (possibly finite) chain
$\overline{V}=V_0,V_1,V_2,\dots$, define $R_{\overline{V}}$ and
$\mathit{succ}^a_{\overline{V}}$ just as they are defined for
sequences of states, but instead of checking that an element at index
$i$ is a call state (resp., a return state or an internal state) by
checking that it is of the form $(-,\call)$ (resp., $(-,\ret)$ or
$(-,\inter)$), do so by checking that $\call\in V_i$ (resp., $\ret\in
V_i$ or $\inter\in V_i$). This is well defined, thanks to axiom
\axiom{C1}.

An infinite chain $\overline{V}=V_0,V_1,\dots$ of atoms is
\emph{acceptable} if for all $i$, 
\begin{itemize}
\item whenever $\psi_1\Until\psi_2\in V_i$, then there exists
$i\le j\le |\sigma|$ such that $\psi_2\in V_j$ and $\psi_1\in
V_i,\dots,V_{j-1}$;
\item whenever $\Next^a\psi\in V_i$, then $\psi\in
V_{\mathit{succ}^a_{\overline{V}}(i)}$;
\item whenever $\psi_1\Until^a\psi_2\in V_i$, then there exists
$i_0,i_1,\dots,i_k$ (with $i_0=i$)
such that $\mathit{succ}^a_{\overline{V}}(i_j)=i_{j+1}\ne\bot$ (for $j=0,\dots,k-1$),
$\psi_2\in V_{i_k}$,  and $\psi_1\in V_{i_j}$ (for $j=0,\dots,k-1$). 
\end{itemize}
The following lemma isolates the properties needed to finish the proof
of completeness.
\begin{lemma}\label{l:2}
\begin{enumerate}
\item 
For all $\Next\psi\in\FL(\phi)$ and atoms $V$, $\Next\psi\in 
V$ if and
only if for all $W\in V\mathord{\lr}$, $\psi\in W$
\item 
For all $\GNext\psi\in\FL(\phi)$ and atoms $V$,
$\GNext\psi\in V$ if and only if there exists
$W\in V\mathord{\lr}$ such that $\psi\in W$.  
\item 
For all $\psi_1\Until\psi_2\in\FL(\phi)$ and atoms $V_0$,
$\psi_1\Until\psi_2\in V_0$ if and only if there exists a finite chain 
$V_0,V_1,\dots,V_k$ such that $\psi_1\in
V_0,\dots,V_{k-1}$ and $\psi_2\in V_k$.  
\item For all $\Next^a\psi\in \FL(\phi)$ and atoms $V_0$,
$\Next^a\psi\in V_0$ if and only if for all finite chains
$\overline{V}$ of atoms $V_0,V_1,\dots,V_k$ such that
$\mathit{succ}^a_{\overline{V}}(0)=k$, $\phi\in V_k$. 
\item 
For all $\psi_1\Until^a\psi_2\in\FL(\phi)$ and atoms $V_0$,
$\psi_1\Until\psi_2\in V_0$ if and only if there exists a finite chain
$\overline{V}$ of atoms
$V_0,V_1,\dots,V_k$ and indices $i_0,\dots,i_j<k$ such that
$\mathit{succ}^a_{\overline{V}}(i_l)=i_{l+1}$ (for $l=0,\dots,j-1$),
$\mathit{succ}^a_{\overline{V}}(i_j)=k$, $\psi_1\in
V_{i_0},\dots,V_{i_j}$ and $\psi_2\in V_k$.  
\item Every finite chain of atoms is extensible to an
acceptable chain. 
\end{enumerate}
\end{lemma}
I leave the proof as an exercise; it follows the structure of the
proof of Lemma~\ref{l:1} quite closely, despite requiring a slightly
more involved argument for parts (d) and (e), as one would expect.

So there we are: a sound and complete axiomatization for (an important
fragment of) \caret{}. We get the usual benefits from it, namely, the
possibility of reasoning purely deductively about structured temporal
structures, and this gives an alternative to model-checking for
proving properties of programs. I do not know, at this point, whether the 
decision problem for the logic is decidable, and so reasoning
deductively may not be feasible.  One problem with the axiomatization
$\AXcr$ is that it is not very nice. In fact, axioms
\axiom{C5} and \axiom{C6} are downright ugly. I believe this is
difficult to avoid. Since \caret{} does not have operators for
counting, the axioms must keep count the hard way---listing all
possibilities---to match returns to calls. It may still be possible,
however, to develop alternate axiomatizations more suited to using
\caret{} as a program logic. That remains to be seen.

\end{document}